\documentclass[aps,prl,twocolumn,groupedaddress]{revtex4}
\usepackage{amssymb}
\usepackage{graphicx}
\usepackage{amssymb}
\usepackage{graphicx}

\begin{document}

\title{Universal doping dependence of the ground state staggered
magnetization in cuprates}
\author{ Rinat Ofer$^{1}$, Amit Keren$^{1}$, Omar Chmaissem$^{2}$, and Alex
Amato$^{3}$} \affiliation{$^{1}$Physics Department, Technion-Israel
Institute of Technology, Haifa 32000, Israel}
\affiliation{{\normalsize {$^{2}$}}Materials Science Division,
Argonne National Laboratory, 9700 S. Cass Avenue, Argonne, IL 60439,
USA} \affiliation{$^{3}$Paul Scherrer Institute, CH 5232 Villigen
PSI, Switzerland}
\date{\today }

\begin{abstract}
Using muon spin rotation we determine the zero temperature staggered
antiferromagnetic order parameter $M_{0}$ versus hole doping measured from
optimum $\Delta p_{m}$, in the (Ca$_{x}$La$_{1-x}$)(Ba$_{1.75-x}$La$%
_{0.25+x} $)Cu$_{3}$O$_{y}$ system. In this system the maximum
$T_{c}$ and the superexchange $J$ vary by 30\% between families
($x$). $M_{0}(x,\Delta p_{m}) $ is found to be $x$-independent.
Using neutron diffraction we also determine the lattice parameters
variations for all $x$ and doping. The oxygen buckling angle is
found to change with $x$, implying a change in the holes kinetic
energy. We discuss the surprising insensitivity of $M_{0}(x,\Delta
p_{m})$ to the kinetic energy variations in the framework of the
$t$-$J$ model.
\end{abstract}

\maketitle

It is widely agreed that the antiferromagnetic (AFM) phase of the cuprates
should be addressed as a doped Mott insulator, where holes are moving on a
2D AFM\ background \cite{DagottoRMP94}. This scenario is described by the $t$%
-$J$ model Hamiltonian where $t$ and $t^{\prime }$ are the near and
next-near neighbor hoppings, respectively, and $J$ is the Heisenberg
superexchange. Above some critical doping the zero temperature\ staggered
AFM order parameter $M_{0}$ is destroyed, and the cuprates enter a glassy,
phase separated, state. Since the glassy state precedes superconductivity,
understanding this transition is crucial to understanding the cuprates.
Particularly interesting is the doping dependence of $M_{0}$ and its
variations with the different energy scales. These variations were
calculated theoretically \cite{MvsP} but not measured in a controlled
manner. Such measurements could shed light on the effective Hamiltonian
governing the holes at $T\rightarrow 0$ in the underdoped region. While $J$
can be measured relatively simply with neutron or Raman scattering on a
single crystal, it is very difficult to determine $t$ experimentally.

In this work we determine $t/J$ from their lattice parameter dependence,
including the buckling angle, for different cuprate families ($x$) of the (Ca%
$_{x}$La$_{1-x}$)(Ba$_{1.75-x}$La$_{0.25+x}$)Cu$_{3}$O$_{y}$ (CLBLCO)
system, where the maximum $T_{c}$ ($T_{c}^{max}$) and $J$ varies by about $%
30 $\% between families \cite{OferPRB}. This is done by Rietvelt
refinement of neutron diffraction. We also determine the doping
dependence of $M_{0}$, which is expected to depend on $t/J$. Zero
field muon spin rotation ($\mu$SR) is employed for this purpose. Our
main finding is that the doping dependence of $M_{0}$ is universal
despite the fact that $t/J$ varies between families.

\begin{figure}[t]
\begin{center}
\includegraphics[width=9cm]{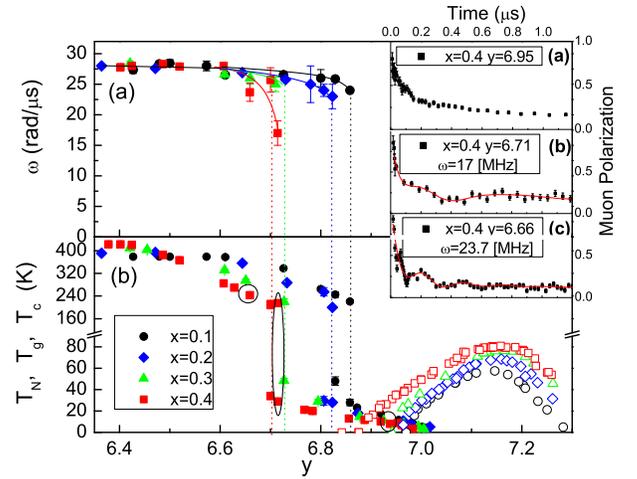}
\end{center}
\caption{(a) The zero temperature muon oscillation angular frequency
$\protect\omega $ as a function of the chemical doping $y$ for all
four of the
(Ca$_{x}$La$_{1-x}$)(Ba$_{1.75-x}$La$_{0.25+x}$)Cu$_{3}$O$_{y}$
(CLBLCO) families. The antiferromagnetic zero temperature order
parameter $M_{0}$ is proportional to $\protect\omega $. (b) The
CLBLCO phase diagram from \protect\cite{OferPRB} including magnetic
(close symbols) and superconducting (open symbols) critical
temperatures. The insets show raw muon polarization data and fits to
Eq. \protect\ref{oscillation fit} for three samples from the $x=0.4$
family marked in panel b.} \label{wvsy}
\end{figure}

We chose to work with the CLBLCO system for several reasons. This is a high
temperature superconductor (HTSC) with YBa$_{2}$Cu$_{3}$O$_{y}$ (YBCO)
structure. The family index $x$ varies in the range $0.1\leq x\leq 0.4$.
Doping is possible all the way to the overdoped regime. All compounds are
tetragonal and there is no chain ordering as in YBCO \cite{Goldschmidt}. As
we show below, there are minimal structural differences between the
families. In addition, the level of disorder as detected by Ca NMR~\cite%
{MarchandThesis} and Cu\ NQR \cite{KerenToBe} is identical for the different
families. The phase diagram is presented in Fig.~\ref{wvsy}(b) showing the
antiferromagnetic N\'{e}el temperature $T_{N}$, the spin glass temperature $%
T_{g}$ where islands of spins freeze, and the superconducting critical
temperature $T_{c}$; note the axis breaker. In this phase diagram $T_{N}$
\cite{OferPRB} and $T_{g}$ \cite{KanigelPRL02} were measured by $\mu $SR,
and $T_{c}$ was measured by resistivity \cite{Goldschmidt}. The spin glass
phase penetrates into the superconducting phase. It also slightly penetrates
into the N\'{e}el phase in the sense that a first transition, to long range
order, takes place near $200$~K, and a second transition, with additional
spontaneous fields, takes place near $10$~K. Each transition is a continuous
function of doping.

\begin{figure}[t]
\begin{center}
\includegraphics {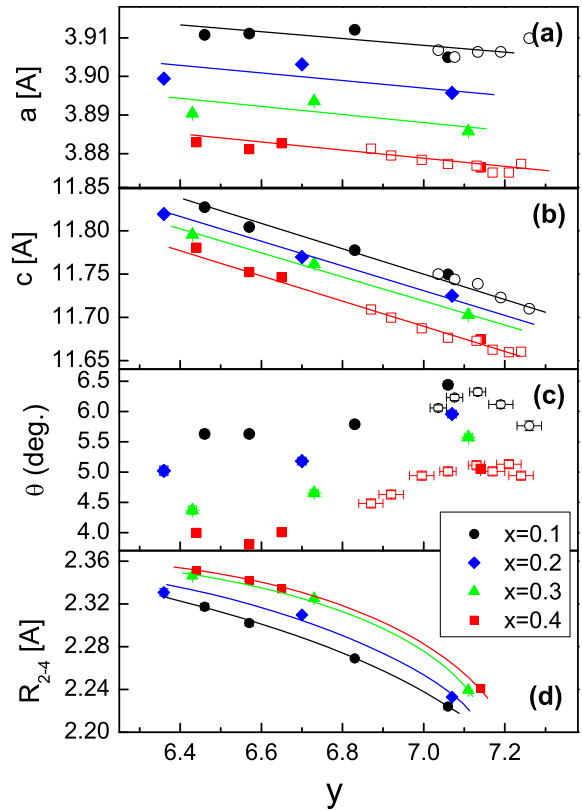}
\end{center}
\caption{The parameters extracted from a neutron diffraction experiment as a
function of oxygen doping for the four families of CLBLCO. (a) The lattice
parameter $a$. (b) The lattice parameter $c$. (c) $\protect\theta $ - the
buckling angle between the copper and oxygen in the plane. (d) $R_{24}$ -
The distance between the in-plane copper and the apical oxygen. The empty
symbols are measurements taken from \protect\cite{Chmaissem}. The lines are
guides to the eye.}
\label{neutrons}
\end{figure}

The neutron powder diffraction experiments were performed at the Special
Environment Powder Diffractometer at Argonne's Intense Pulsed Neutron Source
(see Ref \cite{Chmaissem} for more details). Fig.~\ref{neutrons} shows a
summary of the lattice parameters. The empty symbols represent data taken
from Ref.~\cite{Chmaissem}. All the parameters are family-dependent,
however, not to the same extent. The lattice parameters $a$ and $c$,
depicted in Fig.~\ref{neutrons}(a) and (b), change by up to about $0.5\%$
between the two extreme families ($x=0.1$ and $x=0.4$). The in-plane Cu-O-Cu
buckling angle is shown in Fig.~\ref{neutrons}(c). This angle is non-zero
since the oxygen is slightly out of the Cu plane and closer to the Y site of
the YBCO structure. The buckling angle shows strong variation between the
families; there is about a $30\%$ change from the $x=0.1$ family to $x=0.4$.
This change is expected since as $x$ increases, a positive charge is moving
from the Y to the Ba site of the YBCO structure, pulling the oxygen toward
the plane and flattening the Cu-O-Cu bond.

We believe that this property is the main cause for the different $J$ and
therefore different $T_{c}^{max}$ between the CLBLCO families \cite{OferPRB}%
. Nevertheless, we note that Pavarini \emph{et al. }\cite{Pavarini} showed
that in many cuprates families $T_{c}^{max}$ scales with $t^{\prime }/t$. $%
t^{\prime }$ is controlled by the hybridization of the Cu 4s with the apical
oxygen 2p$_{z}$, hence $T_{c}^{max}$ scales with the distance $R_{24}$
between the in-plane copper and the apical oxygen. In Fig.~\ref{neutrons}(d)
we show $R_{24}$ for our CLBLCO samples. Our results also support
qualitatively Pavarini's conclusion.
\begin{figure}[t]
\begin{center}
\includegraphics {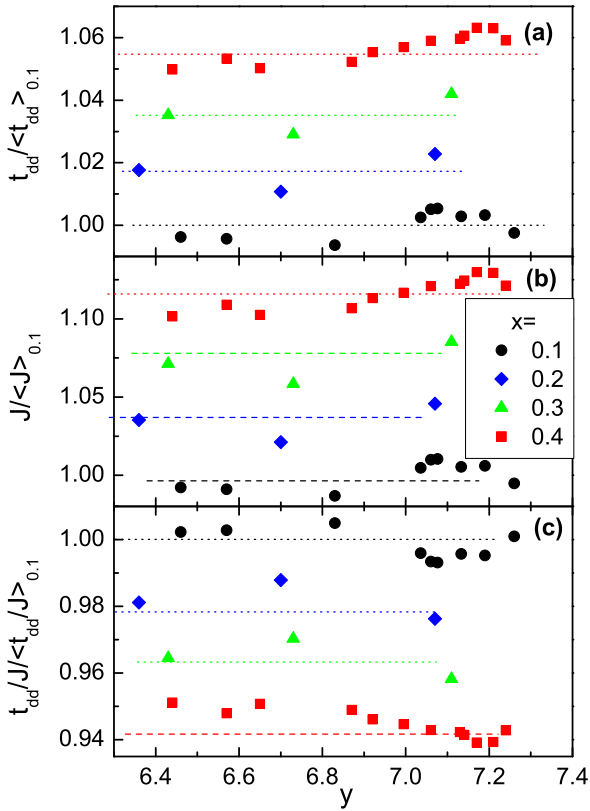}
\end{center}
\caption{(a) The hopping rate $t_{dd}$, and (b) the superexchange coupling $%
J $ calculated from Eq. (\protect\ref{t}) and (\protect\ref{J}), using the
parameters shown in Fig.~\protect\ref{neutrons}. (c) The ratio $t_{dd}/J$ as
a function of doping. The dotted lines are guides to the eye. All data sets
are normalized to the $x=0.1$ familiy.}
\label{tJparam}
\end{figure}

From the lattice parameters and the buckling angle it is possible
to construct the $t/J$ ratio assuming that the Hubbard $U$ and the
charge transfer energy $\Delta $ are family-independent. The basic
quantity is the
hopping integral $t_{pd}$ between a Cu $3d_{x^{2}-y^{2}}$ and O $2p$ \cite%
{ZaaneannCJP}. This hopping integral is proportional to bond length to the
power -3.5 \cite{Harrison}. The hopping from the O $2p$ to the next Cu $%
3d_{x^{2}-y^{2}}$ involves again the bond length and cosine of the angle.
Thus, the Cu to Cu hopping depends on $a$ and $\theta $ as
\begin{equation}
t_{dd}\propto \frac{\cos \theta }{a^{7}}.  \label{t}
\end{equation}%
$J$ is proportional to $t_{dd}^{2}$, hence,
\begin{equation}
J\propto \frac{\cos ^{2}\theta }{a^{14}}.  \label{J}
\end{equation}%
Estimates of the $t_{dd}$ and $J$, normalized to the averaged values of the $%
x=0.1$ family, $\left\langle t_{dd}\right\rangle _{0.1}$ and $\left\langle
J\right\rangle _{0.1}$ are presented in Fig.~\ref{tJparam}(a) and (b).
Although there is a variation in $t$ and $J$ within each family, the
variation is much larger between the families. $J$ increases with increasing
$x$, in qualitative agreement with experimental determination of $J$ \cite%
{OferPRB}. In Fig.~\ref{tJparam}(c) we show the normalized ratio
$t/J$ for the four CLBLCO families. There is about $5\%$
difference between the two extreme families. But we stress that
this determination of $t/J$ is only an estimate, used in practice
to set the oxygen level spacing between samples in the $\mu $SR
experiment. More accurate calculations of $t/J$ are in progress
and preliminary data indicate that $t/J$ varies by more than 10\%
between families \cite{Marie}.

Next we determine the doping dependence of the order parameter using
zero field $\mu $SR. The experiments were done on the GPS beam line
at the Paul Scherrer Institute, Switzerland. The muon oscillation
angular frequency $\omega $ is proportional to the local magnetic
field it experiences. Therefore, it can be used to determine the
staggered magnetization $M$. Typical muon polarization curves at
$T=5$~K are presented in the insets of Fig.~\ref{wvsy} for three
samples from the $x=0.4$ family on the border between N\'{e}el and
glass order. These three samples are marked on the phase diagram in Fig.~\ref%
{wvsy}(b). More raw data can be found in Ref.~\cite{OferPRB}. The sample in
inset (a) is in the spin glass phase; it has no long-range magnetic order
and hence has no oscillations. The sample in inset (c) is in the
antiferromagnetic phase, and so it has strong oscillations at low
temperatures. Finally, the sample in inset (b) is an example of an
intermediate sample and thus has weaker oscillations.

The best fit of the polarization is achieved with the function
\begin{equation}
P(t)=\sum_{i=1}^{3}A_{i}\exp (-\lambda _{i}t)\cos (\omega _{i}t)
\label{oscillation fit}
\end{equation}%
with $\omega _{3}=0$; the fit is shown in insets (b) and (c) of Fig.~\ref%
{wvsy} by the solid line. The reason for multiple frequencies is
that the muons stop at different sites in the unit cell. The order
parameter extracted from the high angular frequency, around a few
tens of of radians per micro-second, is known to agree with neutron
scattering experiments \cite{KerenPRB93}. The lower angular
frequency is believed to emerge from metastable muon sites and is
not used for further analysis.

The muon polarization was measured at low temperatures, typically
from $5$~K to $200$~K, and the oscillation for $T\rightarrow 0$ was
extracted from extrapolation. Fig.~\ref{wvsy}(a) shows a summary of
the oscillation angular frequency $\omega (T\rightarrow 0)$ as a
function of the chemical doping $y$ for all four CLBLCO families. In
this plot the AFM critical doping, where the oscillations disappear,
is different for each family. Not surprisingly, this is the same
oxygen doping where the N\'{e}el order is replaced by the spin glass
phase in the diagram. However, the chemical doping is different from
the mobile hole doping $p_{m}$, and a rescaling of the doping axis
is required.
\begin{figure}[t]
\begin{center}
\includegraphics {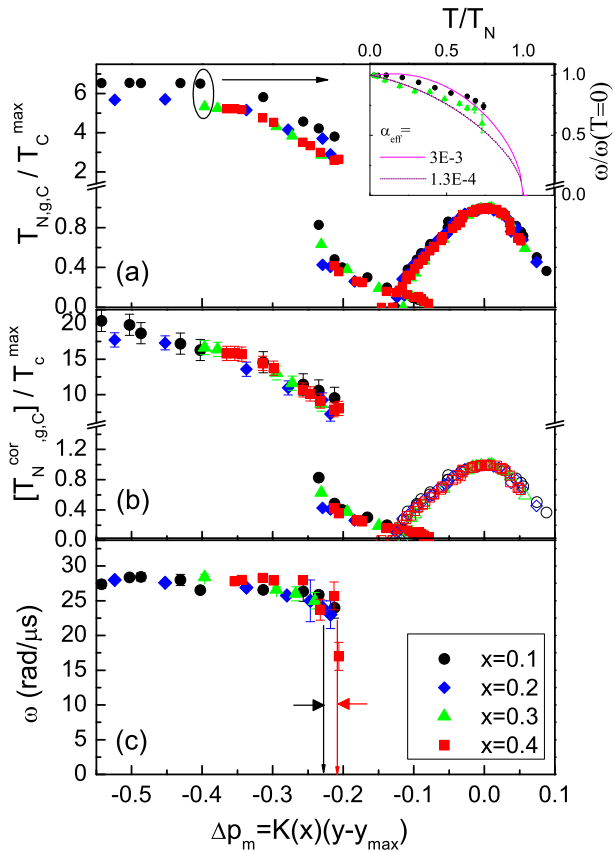}
\end{center}
\caption{(a) The CLBLCO phase diagram after rescaling: For each
family ($x$), the critical temperatures are normalized by $T_{c}$ at
optimal doping, and $y$ is replaced by $\Delta p_{m}$ (see text for
details). (b) The same as (a) but the N\'{e}el temperature is
corrected for anisotropy contribution so it is the same as $J$ for
the parent compound (see text). (c) The zero temperature muon
oscillation angular frequency as a function of $\Delta p_{m}$ for
all four CLBLCO families; an equivalent to the staggered
magnetization $M_{0}$ versus mobile hole density. The arrows show
the expected variation of the critical doping from a $5$\% variation
in $t/J$.} \label{wvsDp}
\end{figure}

The scaling is designed so that the $T_{c}$ domes of all families,
normalized by $T_{c}^{max}$, will collapse on to a single dome. For this
purpose the mobile hole parameter measured from optimum $\Delta p_{m}$ is
defined by $\Delta p_{m}=K(x)\cdot (y-y_{max})$, where $y_{max}$ is the
optimal oxygen doping, and $K(x)$ is a family-dependent scaling parameter. $K
$ should be thought of as doping efficiency parameter connecting oxygen
level to mobile holes in the CuO$_{2}$ planes. The best scaling was found
using $K=0.76,0.67,0.54,0.47$ for $x=0.1\ldots 0.4$, respectively, and is
shown in Fig.~\ref{wvsDp}(a). The errors of $K$ are discussed below. Despite
the fact that $K(x)$ was chosen to scale the $T_{c}$ domes, by the same
token other critical temperatures scale as well. Fig.~\ref{wvsDp}(a) also
shows, $T_{g}/T_{c}^{max}$ and $T_{N}/T_{c}^{max}$, for all families, as a
function of $\Delta p_{m}$. The $T_{g}/T_{c}^{max}$ curves of all families
collapse onto each other and $T_{N}/T_{c}^{max}$ for the $x=0.2$ to $0.4$
families collapse onto each other. The reason $T_{N}$ of the $x=0.1$ family
is not in-line with the others is due to interactions between planes.

The N\'{e}el temperature stems from the three dimensional interaction, and
it is a function of not only the in-plane $J$, but also interplane couplings
$J_{\bot }$ and other anisotropies. In the Heisenberg model $%
T_{N}=Jt_{N}(\alpha _{eff})$ where the effective anisotropy $\alpha
_{eff}$ is mainly set by $J_{\bot }/J$, and $t_{N}(\alpha _{eff})$
is a known logarithmic function of the anisotropy \cite{SBMF}. We
determine $\alpha _{eff}$ and extract $J$ from $T_{N}$. This is done
by measuring the temperature dependence of the muon rotation angular
frequency $\omega (T)$, which is proportional to order parameter
$M(T)$. The function $M(T/T_{N})/M_{0}$
depends only on $\alpha _{eff}$. The inset of Fig.~\ref{wvsDp}(a) shows $%
\omega (T/T_{N})/\omega (T\rightarrow 0)$ for two samples marked in Fig.~\ref%
{wvsDp}(a), and a fit to the predicted behavior given in Refs.~\cite{SBMF}
and \cite{OferPRB}. The $x=0.1$ sample clearly has a bigger $\alpha _{eff}$
than the $x=0.3$ sample, and is more 3D like. Using this method we
determined $\alpha _{eff}$ for all samples with N\'{e}el order and defined
the quantity $T_{N}^{cor}=T_{N}/t_{N}(\alpha _{eff})$ for these samples \cite%
{OferPRB}. For zero doping $T_{N}^{cor}=J$ . When the system is doped, $%
T_{N} $ is also affected by hopping and $T_{N}^{cor}=J$ is no longer valid. $%
T_{N}^{cor}$ replaces $T_{N}$ in Fig.~\ref{wvsDp}(b) which otherwise is the
same as Fig.~\ref{wvsDp}(a). After this replacement the entire phase diagram
scales to a single unified curve, indicating that $T_{c}^{max}\propto J$ and
that a single energy scale controls magnetism and superconductivity.

Fig.~\ref{wvsDp}(c) shows $\omega (T\rightarrow 0)$ as a function of
$\Delta p_{m}$ for each family. The scalability of the phase
diagram, as explained above, suggests that $\Delta p_{m}$ is a
parameter proportional to the mobile hole density variation. Hence
Fig.~\ref{wvsDp}(a) is equivalent to a plot of the AFM order
parameter at zero temperature as a function of mobile hole density.
This plot shows that the order parameter is universal for all
families, and in particular the AFM critical doping is
family-independent. To demonstrate this point we show, using the two
arrows in Fig.~\ref{wvsDp}(c), what should have been the difference
in the critical doping had it been proportional to $t/J$, and
changed between the $x=0.4$ and $x=0.1$ by 5\% (of $0.3$) as
indicated in Fig. \ref{tJparam}(c). Thus, we conclude that
$M_{0}(x,\Delta p_{m})$ is $x$-independent, hence independent of
$t/J$.

The above conclusion could, \emph{a priori}, depend on the choice of
the $K$'s and $y_{max}$. A different set of $K$'s or $y_{max}$ would
shift the magnetic critical doping with respect to each other.
However, it will also shift the normalized $T_{c}$ domes, $T_{g}$
line, and $T_{N}^{cor}$ line with respect to each other. We have
attempted to use a different set of $K$'s and $y_{max}$'s, which
will not noticeably destroy the scaling of the normalized critical
temperatures. We could not generate a variation of more than 2\% in
the $M_{0}$ critical doping. In other words, the different sets of
$K$'s and $y_{max}$'s always kept the critical doping well between
the two vertical arrows in Fig.~\ref{wvsDp}(c).

This surprising result could be discussed using three scenarios: (I)
That different values of $t/J$ correspond to different values of
$K$, namely, the changes in $t/J$ are cancelled out by the re-scaled
doping axis. However, there is no theoretical backing for his
scenario. (II) It may be that at low temperatures the effective
Hamiltonian is given by a $t$-$J$ model but with an effective $t$
that is proportional to $J$. Indeed, there are indications that for
small hole doping in an antiferromagnet the bandwidth of the hole
dispersion is set by $J$ (instead of $t$) \cite{tJ}. However, the
spatial size of each hole quasiparticle (spin--polaron), and thus
the critical doping, does depends on $t/J$. Moreover, it is not
clear how spin-polarons destroy the AFM order and at which doping.
(III) An alternative explanation is that the destruction of the AFM
order parameter is not a result of single holes hopping and should
be described by a completely different effective Hamiltonian;
perhaps hopping of boson pairs \cite{HavilioPRL98}. In this case $t$
should be absorbed into the creation of tightly bound bosons leaving
a prominent energy scale $J$. The proximity of the magnetic critical
doping to superconductivity makes this possibility appealing.

The last possibility could also solve a profound riddle in the study
of the CLBLCO system. This system was found to obey the Uemura
relation $T_{c}\propto n_{s}$ \cite{Uemura}, where $n_{s}$ is the
superconducting carrier density, in both under- and overdoped
regions \cite{KerenSSC03}. At the same time $T_{c}^{max}$ scales
with $J$ as indicated before. Therefore, the Uemura relation should
be rewritten as $T_{c}\propto Jn_{s}$. What is then the role of $t$?
Our finding that the magnetic order parameter versus doping is
universal suggests that even before superconductivity appears, $t$
becomes less relevant. This suggestion does have theoretical support
\cite{KancharlaPRB08}.

To conclude, an estimate of $t$ and $J$ from simple structure
considerations using neutron diffraction shows that the origin of
the different energy scales between the CLBLCO families is mainly
the different buckling angles. The difference in $t/J$ between the
two extreme families is about $5\%$. Although this is not an
accurate way to measure the hopping rate or superexchange coupling,
it does set the scale for the expected variation in the AFM critical
doping. Using $\mu$SR, the AFM order parameter as a function of
oxygen was determined for different families of the CLBLCO system.
We used a scaling transformation to move from oxygenation level to
mobile holes. Our measurements show that, at zero temperature, the
order parameter as a function of mobile holes is independent of
$t/J$ within the required accuracy.

We would like to thank D-G. A. Sawatzky, M-B. Lepetit, A. Auerbach,
and E. Amit for very helpful discussions. We acknowledge financial
support from the Israel Science Foundation, the European Commission
under the 6th Framework Programme, and the Posnansky research fund
in high temperature superconductivity. We are also grateful to the
PSI facilities for high quality muon beams and technical support.
This work was partially supported by the Division of Materials
Science and Engendering of the Office of the Basic Energy Science,
U.S. department of Energy of Science, under Contact No.
DE-AC02-06CH11357, and through the Key Action: Strengthening the
European Research Area, Research Infrastructures. Contract nr:
RII3-CT-2003-505925

\end{document}